\newcommand{\CLASSINPUTtoptextmargin}{19mm}%
\newcommand{\CLASSINPUTbottomtextmargin}{43mm}%
\newcommand{\CLASSINPUTinnersidemargin}{12.9mm}%
\newcommand{\CLASSINPUToutersidemargin}{12.9mm}%
\documentclass[conference,10pt,a4paper]{IEEEtran}%
\usepackage{amsmath}
\usepackage{times}
\usepackage{graphicx}
\usepackage{multirow}
\usepackage[none]{hyphenat}
\makeatletter

\def\@maketitle{\newpage
\bgroup\par\addvspace{0.5\baselineskip}\centering%
\ifCLASSOPTIONtechnote
   {\bfseries\large\@IEEEcompsoconly{\sffamily}\@title\par}\vskip 1.3em{\lineskip .5em\@IEEEcompsoconly{\sffamily}\@author
   \@IEEEspecialpapernotice\par{\@IEEEcompsoconly{\vskip 1.5em\relax
   \@IEEEtitleabstractindextextbox{\@IEEEtitleabstractindextext}\par
   \hfill\@IEEEcompsocdiamondline\hfill\hbox{}\par}}}\relax
\else
   \vskip0.2em{\EuMWtitlesize\ifCLASSOPTIONtransmag\bfseries\LARGE\fi\@IEEEcompsoconly{\sffamily}\@IEEEcompsocconfonly{\normalfont\normalsize\vskip 2\@IEEEnormalsizeunitybaselineskip
   \bfseries\Large}\@title\par}\vskip1.0em\par
   \ifCLASSOPTIONconference%
      {\@IEEEspecialpapernotice\mbox{}\vskip\@IEEEauthorblockconfadjspace%
       \mbox{}\hfill\begin{@IEEEauthorhalign}\@author\end{@IEEEauthorhalign}\hfill\mbox{}\par}\relax
   \else
      \ifCLASSOPTIONpeerreviewca
         {\@IEEEcompsoconly{\sffamily}\@IEEEspecialpapernotice\mbox{}\vskip\@IEEEauthorblockconfadjspace%
          \mbox{}\hfill\begin{@IEEEauthorhalign}\@author\end{@IEEEauthorhalign}\hfill\mbox{}\par
          {\@IEEEcompsoconly{\vskip 1.5em\relax
           \@IEEEtitleabstractindextextbox{\@IEEEtitleabstractindextext}\par\hfill
           \@IEEEcompsocdiamondline\hfill\hbox{}\par}}}\relax
      \else
         \ifCLASSOPTIONtransmag
           {\@IEEEspecialpapernotice\mbox{}\vskip\@IEEEauthorblockconfadjspace%
            \mbox{}\hfill\begin{@IEEEauthorhalign}\@author\end{@IEEEauthorhalign}\hfill\mbox{}\par
           {\vspace{0.5\baselineskip}\relax\@IEEEtitleabstractindextextbox{\@IEEEtitleabstractindextext}\vspace{-1\baselineskip}\par}}\relax
         \else
           {\lineskip.5em\@IEEEcompsoconly{\sffamily}\sublargesize\@author\@IEEEspecialpapernotice\par
           {\@IEEEcompsoconly{\vskip 1.5em\relax
            \@IEEEtitleabstractindextextbox{\@IEEEtitleabstractindextext}\par\hfill
            \@IEEEcompsocdiamondline\hfill\hbox{}\par}}}\relax
         \fi
      \fi
   \fi
\fi\par\addvspace{0.0\baselineskip}\egroup}

\def\EuMWtitlesize{\@setfontsize{\EuMWtitlesize}{24}{24pt}}
\def\EuMWauthorsize{\@setfontsize{\EuMWauthorsize}{11}{11pt}}
\def\EuMWaffilsize{\@setfontsize{\EuMWaffilsize}{10}{10pt}}
\def\EuMWcaptionsize{\@setfontsize{\EuMWcaptionsize}{9}{10pt}}
\def\EuMWbibsize{\@setfontsize{\EuMWbibsize}{8}{10pt}}

\def\@IEEEauthorblockNstyle{\EuMWauthorsize\@IEEEcompsocnotconfonly{\sffamily}\@IEEEcompsocconfonly{\large}}
\def\@IEEEauthorblockAstyle{\EuMWaffilsize\@IEEEcompsocnotconfonly{\sffamily}\@IEEEcompsocconfonly{\itshape}\@IEEEcompsocconfonly{\large}}
\def\@IEEEauthordefaulttextstyle{\EuMWauthorsize\@IEEEcompsocnotconfonly{\sffamily}\sublargesize}

\def\thebibliography#1{\section*{\refname}%
    \addcontentsline{toc}{section}{\refname}%
    \EuMWbibsize\@IEEEcompsocconfonly{\small}\vskip 0.3\baselineskip plus 0.1\baselineskip minus 0.1\baselineskip
    \list{\@biblabel{\@arabic\c@enumiv}}%
    {\settowidth\labelwidth{\@biblabel{#1}}%
    \leftmargin\labelwidth
    \advance\leftmargin\labelsep\relax
    \itemsep \IEEEbibitemsep\relax
    \usecounter{enumiv}%
    \let\p@enumiv\@empty
    \renewcommand\theenumiv{\@arabic\c@enumiv}}%
    \let\@IEEElatexbibitem\bibitem%
    \def\bibitem{\@IEEEbibitemprefix\@IEEElatexbibitem}%
\def\newblock{\hskip .11em plus .33em minus .07em}%
\ifCLASSOPTIONtechnote\sloppy\clubpenalty4000\widowpenalty4000\interlinepenalty100%
\else\sloppy\clubpenalty4000\widowpenalty4000\interlinepenalty500\fi%
    \sfcode`\.=1000\relax}

\long\def\@makecaption#1#2{%
\ifx\@captype\@IEEEtablestring%
\par\@IEEEtabletopskipstrut
\else
\@IEEEfigurecaptionsepspace
\fi
\setbox\@tempboxa\hbox{\normalfont\footnotesize {#1.}\nobreakspace\nobreakspace #2}%
\ifdim \wd\@tempboxa >\hsize%
\setbox\@tempboxa\hbox{\normalfont\footnotesize {#1.}\nobreakspace\nobreakspace}%
\parbox[t]{\hsize}{\normalfont\footnotesize\noindent\unhbox\@tempboxa#2}%
\else
\ifCLASSOPTIONconference \hbox to\hsize{\normalfont\footnotesize\hfil\box\@tempboxa\hfil}%
\else \hbox to\hsize{\normalfont\footnotesize\box\@tempboxa\hfil}%
\fi\fi
\ifx\@captype\@IEEEtablestring%
\@IEEEtablecaptionsepspace
\else
\fi}

\newlength\tablecaptiontotableskip
\newlength\figuretocaptionskip
\def\@IEEEfigurecaptionsepspace{\vskip\figuretocaptionskip\relax}%
\def\@IEEEtablecaptionsepspace{\vskip\tablecaptiontotableskip\relax}%

\def\abstract{\normalfont%
\@IEEEabskeysecsize\bfseries\textit{\abstractname}\,\bfseries\textit{---}\,%
\@IEEEgobbleleadPARNLSP}%

\def\IEEEkeywords{\normalfont%
\@IEEEabskeysecsize\bfseries\textit{\IEEEkeywordsname}\,\bfseries\textit{---}\,%
\@IEEEgobbleleadPARNLSP}%
\def\endIEEEkeywords{\relax\vspace{0.67ex}%
\par\if@twocolumn\else\endquotation\fi%
\normalsize\normalfont}%

\DeclareRobustCommand*{\EuMWauthorrefmark}[1]{\raisebox{0pt}[0pt][0pt]{\textsuperscript{\footnotesize{#1}}}}%
\def\@IEEEauthorblockNtopspace{0ex}
\def\@IEEEauthorblockAtopspace{1mm}
\def\tablename{Table}
\def\thetable{\arabic{table}}
\def\IEEEkeywordsname{Keywords}
\def\subsubsection{\@startsection{subsubsection}{3}{\z@}{1.5ex plus 1.5ex minus 0.5ex}%
{0.7ex plus .5ex minus 0ex}{\normalfont\normalsize\itshape}}%
\newlength{\CPheadmatchindent}%
\def\@seccntformat#1{\hbox to\CPheadmatchindent{\csname the#1dis\endcsname}\hskip 0.1em \relax}
\IEEEilabelindentA \parindent
\IEEEilabelindent \IEEEilabelindentA
\IEEEelabelindent \parindent
\IEEEdlabelindent \parindent
\IEEElabelindent \parindent
\makeatother

\usepackage{cite}
\usepackage{graphicx}
\usepackage{array}
\usepackage{enumerate}
\usepackage{amsbsy}
\usepackage{amssymb}
\usepackage{amsthm}
\usepackage{amscd}
\usepackage{amsmath}
\usepackage{algpseudocode}
\usepackage{amsthm}
\usepackage{dsfont}
\usepackage{algorithm}
\usepackage{thmtools}
\usepackage{amssymb}
\usepackage[dvipsnames]{xcolor}

\def\mbf{\mathbf{f}}

\def\mbx{\mathbf{x}}
\def\mby{\mathbf{y}}

\def\mbB{\mathbf{B}}
\def\mbC{\mathbf{C}}

\def\mbG{\mathbf{G}}

\def\mbI{\mathbf{I}}

\def\bzero{\boldsymbol{0}}

\def\Diag#1{\mathrm{Diag}\left(#1\right)}

\def\arg#1{\mathrm{arg}\left(#1\right)}

\newcommand{\zinv}[1][]{z^{\ifx&#1&-1\else-#1\fi}}

\theoremstyle{definition}

\newtheorem{remark}{Remark}

\algnewcommand\algorithmicinput{\textbf{Input:}}
\algnewcommand\Input{\item[\algorithmicinput]}
\algnewcommand\algorithmicoutput{\textbf{Output:}}
\algnewcommand\Output{\item[\algorithmicoutput]}
\algnewcommand\algorithmicinit{\textbf{Initialize:}}
\algnewcommand\Init{\item[\algorithmicinit]}

\def\j{\mathrm{j}}

 \def\rect#1{\mathrm{rect}\left(#1\right)}
\def\T{\top}
\def\H{\mathrm{H}}
\usepackage{tikz}
\usepackage{flushend}
\usepackage{caption}
\usepackage{subcaption}
\IEEEoverridecommandlockouts

\begin{document}
\title{Mutual Interference Mitigation in PMCW Automotive Radar}
\author{%
\IEEEauthorblockN{%
Zahra Esmaeilbeig\EuMWauthorrefmark{\#1$\dagger$}, 
Arindam Bose\EuMWauthorrefmark{*2$\dagger$}, 
Mojtaba Soltanalian\EuMWauthorrefmark{\#3}
}
\IEEEauthorblockA{%
\EuMWauthorrefmark{\#}ECE Department, University of Illinois Chicago; 
\EuMWauthorrefmark{*}KMB Telematics, Inc. \\
\{\EuMWauthorrefmark{1}zesmae2, \EuMWauthorrefmark{3}msol\}@uic.edu, \EuMWauthorrefmark{2}abose@kmb.ac} 
\thanks{\EuMWauthorrefmark{$\dagger$}First two authors have equal contributions. This work was sponsored by the National Science Foundation Grant ECCS-1809225.}} 
\maketitle
\begin{abstract}
This paper addresses the challenge of mutual interference in phase-modulated continuous wave (PMCW) millimeter-wave (mmWave) automotive radar systems. The increasing demand for advanced driver assistance systems (ADAS) has led to a proliferation of vehicles equipped with mmWave radar systems that operate in the same frequency band, resulting in mutual interference that can degrade radar performance creating safety hazards. We consider scenarios involving two similar PMCW radar systems and propose an effective technique for a cooperative design of transmit waveforms such that the mutual interference between them is minimized. The proposed approach is numerically evaluated via simulations of a mmWave automotive radar system. The results demonstrate that the proposed technique notably reduces mutual interference and enhances radar detection performance while imposing very little computational cost and a negligible impact on existing infrastructure in practical automotive radar systems.
\end{abstract}

\section{Introduction}
Millimeter-wave (mmWave) automotive radar systems have gained significant attention due to their accurate object detection capabilities in challenging environments. Compared to cameras and Lidar, mmWave radar systems excel in heavy rain, fog, snow, and smoke \cite{8792451, elbir2022twenty}. Operating within the 77 GHz to 81 GHz frequency range, these systems utilize high-frequency continuous waves (CW) for object detection. However, their poor angular resolution limits the detection of fine spatial details. Multiple-input multiple-output (MIMO) technology can improve the resolution, but it also introduces mutual interference challenges.

Mutual interference arises in MIMO radar systems when multiple transmitters operate in close proximity, leading to increased noise floor and reduced detection accuracy and reliability. Advanced signal processing techniques such as digital beamforming and adaptive filtering can mitigate this issue \cite{6854614}. As the number of radar systems in vehicles grows and the mmWave frequency band becomes more congested, mutual interference becomes increasingly problematic \cite{alland2019interference}. Modulation techniques, such as phase-modulated continuous wave (PMCW) and orthogonal frequency-division multiplexing (OFDM), offer advantages over traditional frequency-modulated continuous wave (FMCW), but they require higher sampling rates and sophisticated transceiver hardware \cite{8828004}.

Research in mutual interference mitigation has focused on waveform design for FMCW radar systems \cite{9525348, bose2022waveform, 8207776,1412035, 8378779}. 
Adaptive waveforms applied in slow-time or fast-time signals have been proposed, including slow-time coded waveforms, fast-generating adaptive slow-time coding schemes, and fast-time coding schemes \cite{4487195}. Pseudo-orthogonal noise waveforms and specialized slow-time waveforms like the golden code and the linear frequency modulated CAZAC code have also been explored \cite{8207776, 1412035, 8378779}.

This paper focuses on mitigating mutual interference among PMCW radars. While studies have addressed mutual interference between FMCW and PMCW radars (see \cite{beise2018virtual, yildirim2019impact, 7944482, xu2019doppler} and the references within), there is a research gap in waveform design for PMCW systems, especially when radar systems possess similar physical parameters. To address this gap, a novel framework for designing collaborative waveforms is proposed. The framework can handle non-convex objectives, is computationally efficient for practical implementation, and requires minimal modifications to the transceiver infrastructure.
\section{Problem formulation}
\begin{figure}[t]
\centering
	\input{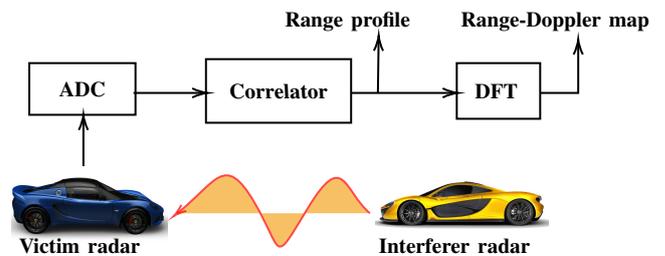}
 \caption{\footnotesize A simplified illustration of the PMCW radar operation and PMCW-PMCW mutual interference between two vehicles.} 
\label{fig_1}
\end{figure}
\subsection{PMCW Signal Model}
In this section, we formulate the PMCW radar model.
We consider two PMCW automotive radar systems, depicted in Fig.~\ref{fig_1}, which are similar and mutually cooperative. These radars operate within the same frequency band. The transmit signal of both PMCW radars in a single \emph{burst} of the signal can be described as
\begin{equation}
 s(t)=\exp(\j (2\pi f_c t+ \phi(t))), \; 0\leq t \leq T,
\end{equation}
where $f_c$ is the carrier frequency, $T$ is the pulse repitition interval (PRI) and $\phi(t)$ is the modulation phase waveform \cite{alland2019interference}.
 If the \emph{chip} duration is $T_c$, then we denote the phase shift of the $k$-th chip by $x_k=e^{\j \phi(t)}$ for $t$ in the interval $k T_c \leq t \leq (k+1)T_c$, resulting in  $s(t)=\sum_{k=0}^{K-1} x_k \exp(\j 2\pi f_c t)\rect{\frac{t-kT_c}{T_c}}$. We assume that $N$ \emph{bursts} of the signal are transmitted in one coherent processing interval (CPI). Therefore, the transmitted signal is\par \noindent \small
\begin{align}\label{eq:12}
S(t)&=\frac{1}{N}\sum_{n=0}^{N-1}s(t-nT)\nonumber\\
&=\frac{1}{N}\sum_{n=0}^{N-1}\sum_{k=0}^{K-1} x_k e^{\j 2\pi f_c t}\rect{\frac{t-kT_c-nT}{T_c}},
\end{align}\normalsize
where $0\leq t \leq NT$, and\par \noindent \small
\begin{equation}
\rect{t}=\begin{cases}
1 & 0 \leq t \leq 1, \\
0 & \text{otherwise}.
\end{cases}
\end{equation}\normalsize
We consider a single target, located at range $R$ and moving with velocity $v$ towards the radar, which reflects back the radar signal. The two-way target propagation delay is $\tau_{_T}(t)=\frac{2(R-vt)}{c}=\gamma_{_{T}}-\frac{2v}{c}t$, where $c$ is the speed of light. The received signal is 
\par \noindent \small
\begin{align}
S_R(t) &= \alpha_T S(t-\tau_{_T}(t))\nonumber\\
&\approx \frac{\alpha_T}{N} e^{\j 2\pi f_c t}e^{-\j 2\pi f_c \gamma_{_T}}e^{\j 2\pi f_c \frac{2v}{c}t}~\times \nonumber\\&\qquad\sum_{n=0}^{N-1}\sum_{k=0}^{K-1} x_{_{k}} \rect{\frac{t-\gamma_{_T}-kT_c-nT}{T_c}},
\end{align}\normalsize
where we assumed $v \ll c$ for the approximation. We assume the cooperative performance of the two radars eliminates the \emph{carrier frequency offset (CFO)} in the receiver. After mixing the received signal $S_R(t)$ with the conjugate of the carrier frequency, we assume the term $e^{-\j 2\pi f_c \gamma_{_T}}$ is absorbed in $\alpha_T$. We denote $f_{_{d,T}}=\frac{2v}{c}f_c$ as the Doppler frequency to obtain
\begin{equation}
 \hat{S}_{R}(t)= \frac{\alpha_T}{N} e^{\j 2\pi f_{_{d,T}}t} \sum_{n=0}^{N-1}\sum_{k=0}^{K-1} x_k \rect{\frac{t-\gamma_{_T}-kT_c-nT}{T_c}}.
\end{equation}

As illustrated in Fig.~\ref{fig_1}, in the ADC, the time is split into fast time $t'$ and slow time index $n$, with time interval $T$ as $t=t'+nT$, $t' \in [0,T)$. In the fast-time, the signal can be sampled with interval $T_c$, i.e., at $t'=mT_c$, to obtain\par\noindent\small
\begin{align}\label{eq:11}
r_{_T}[&m,n]=\hat{S}_{R}(mT_c+nT)\nonumber\\
&=\alpha_T e^{\j 2\pi f_{_{d,T}}(mT_c+nT)}\sum_{k=0}^{K-1} x_{_{k}} \rect{\frac{(m-k)T_c-\gamma_{_T}}{T_c}}\nonumber\\
&= \alpha_T e^{\j 2\pi f_{_{d,T}}(mT_c+nT)}\sum_{k=0}^{K-1} x_{_{k}} \delta_{_{m-\hat{n}_{_T},k}} \nonumber \\
&=\alpha_T e^{\j 2\pi f_{_{d,T}}(mT_c+nT)} x_{_{m-\hat{n}_{_T}}},
\end{align}\normalsize
where $\delta_{i,j}$ is the extension of the Kronecker delta function with
\begin{equation}
\delta_{i,j}=\begin{cases}
1 & i=j, \\
0 & \text{otherwise},
\end{cases}
\end{equation}
and $\hat{n}_{_T}=\lfloor\frac{\gamma_{_T}}{T_c}\rfloor$ is the number of code shifts due to the target at a range $R$.
As illustrated in Fig.~\ref{fig_1}, in the receiver, the discrete signal in \eqref{eq:11} will go through the correlator to yield the range profiles. The correlation between \eqref{eq:11} and \eqref{eq:12}
is 
\begin{equation}
r_{_T}[m,n]=\alpha_T \sum_{k=0}^{K-1} x_{{_k}}^{*} x_{_{k-\hat{n}_{_T}+m}} e^{\j 2\pi f_{_{d,T}}((m+k)T_c+nT)},
\end{equation}
which is the range profile of the target \cite{yildirim2019impact}. The impact of the second radar, acting as an interferer, on the range profile observed in the victim radar is represented in the following model.
\subsection{Mutual Interference Model} 
The victim radar system receives a signal from the interferer radar transmitter that can be falsely interpreted by the receiver as a reflected signal from a target. Such interference when the two radars are transmitting PMCW is referred to as PMCW-PMCW interference~\cite{bordes2019pmcw}.
In this section, the interferer PMCW radar system, transmitting signal with phase code $\mby=[y_{_0},\ldots,y_{_{K-1}}]^{\T}$, similar to \eqref{eq:12} is assumed to interfere with the victim radar transmitting PMCW signal with phase code $\mbx=[x_{_{0}},\ldots,x_{_{K-1}}]^{\T}$. We define the one-way delay associated with the interference as $\tau_{_I}(t)=\frac{(R_{_I}-v_{_I}t)}{c}=\gamma_{_{I}}-\frac{v_{_I}}{c}t$, where $R_{_I}$ is the distance between two radar systems and $v_{_I}$ is the relative velocity between the two. Let $f_{_{d,I}}=\frac{v_{_I}}{c}f_c$ be the Doppler frequency associated with the interference. The interference samples in the receiver of the victim radar are 
\begin{align}
r_{_{I}}[m,n]&= \alpha_{_{I}} \sum_{k=0}^{K-1} x_{{_k}}^{*} y_{_{k-\hat{n}_{_I}+m}}e^{\j 2\pi f_{_{d,I}}((m+k)T_c+nT)}
\end{align}
with $\hat{n}_{_{I}}=\lfloor\frac{\gamma_{_{I}}}{T_c}\rfloor$ being the number of code shifts due to interference. It is worth highlighting that the cooperative performance of the two radars allows us to effectively compensate for the desynchronization and differing PRI between them.\nocite{esmaeilbeig2023moving}
\section{Mutual interference mitigation}
The received signal is formulated as\par\noindent\small
\begin{align}\label{eq:13}
 r&[m,n]= r_{_{T}}[m,n]+r_{_{I}}[m,n]+w[m,n] \nonumber\\
 &= \alpha_T \sum_{k=0}^{K-1} x^{*}_{{_k}} x_{_{k-\hat{n}_{_T}+m}} e^{\j 2\pi f_{_{d,T}}((m+k)T_c+nT)}\nonumber\\&+\alpha_{_{I}} \sum_{k=0}^{K-1} x^{*}_{{_k}} y_{_{k-\hat{n}_{_I}+m}}e^{\j 2\pi f_{_{d,I}}((m+k)T_c+nT)}+w[m,n],
\end{align}\normalsize
where $w[m,n]$ represents the signal-independent disturbance, e.g., the receiver noise. 
In the receiver, \eqref{eq:13} will go through the
Doppler processor by applying the discrete Fourier transform (DFT) to the slow-time samples. As a result, the range-Doppler map is \par\noindent\footnotesize
\begin{align}\label{eq:3}
 &\textrm{RD}[m,p]= \alpha_T D_{N}(\tilde{f}_{_{d,T}}-p/N) \sum_{k=0}^{K-1} x_{_{k}}^{*} x_{_{k-\hat{n}_{_T}+m}} e^{\j 2\pi f_{_{d,T}}(m+k)T_c} \small \nonumber\\
 &+\alpha_{_{I}} D_{N}(\tilde{f}_{_{d,I}}-p/N)\sum_{k=0}^{K-1} x^{*}_{{_k}} y_{_{k-\hat{n}_{_I}+m}}e^{\j 2\pi f_{_{d,I}}(m+k)T_c}+ W[m,p]
\end{align}\normalsize
where $\tilde{f}_{_{d,T}}=f_{_{d,T}}T$, $\tilde{f}_{_{d,I}}=f_{_{d,I}}T$ and $D_n(x)=\frac{sin(n\pi x)}{(\pi x)}$ is the Dirichlet function. A moving target changes the phases of the chips. This phenomenon is indicated by the $e^{\j 2\pi f_{\cdot,\cdot}(m+k)T_c}$ terms in~\eqref{eq:3}. As a result, the received sequence will not be a pure binary sequence. This sensitivity to Doppler-induced phaseshift known as \emph{Doppler intolerance}~\cite{davis2007phase} creates small sidelobes along the range profile as shown in Fig.~\ref{fig:fig2}. It is also readily known that, for PMCW radars, the range-Doppler estimations are not coupled \cite{yildirim2019impact}.

As mentioned in \cite{beise2018virtual}, the typical Doppler frequency is very low compared to the time scale of fast-time processing i.e. $ f_{_{d, I}}\ll 1/T_c $. 
It is evident from \eqref{eq:3}, that the interference is scaled by the cross-correlations in each range bin.
We define 
\begin{equation}\label{eq:4}
r_{xy}^{l}(f)= \sum_{k=0}^{K-1} x^{*}_{{_k}} y_{_{(k+l)\textrm{mod} K}}e^{\j 2\pi kf}
\end{equation}
and $\hat{f}_{_{d,I}}=f_{_{d,I}}T_c$. One can verify from \eqref{eq:3} that 
$|r_{xy}^{l}(\hat{f}_{_{d,I}})|$ for $l=m-\hat{n}_{_I}$ is a dominant interfering term. 
 In order to mitigate the mutual interference between the two radar systems, we propose to suppress the interference power $|r_{xy}^{l}(\hat{f}_{_{d, I}})|^{2}$. 
Hence, we consider the following optimization problem with respect to the two codes $\mbx$ and $\mby$:
\begin{align}
\mathcal{P}_{1}: \underset{\mbx,\mby}{\textrm{minimize}}& \quad |r_{xy}^{l}(f)|^{2}\nonumber \\
\textrm{subject to} & \quad |x_k|=1, \;|y_k|=1, \; \forall \, k.
\end{align}
In practical scenarios, the victim and interferer radar on vehicles have asynchronous transmission and neither $\hat{n}_{_I}$ nor $\hat{f}_{d, I}$ are known. Therefore, we seek to minimize interference on multiple grid points as\par\noindent\small
\begin{align}\label{eq:opt6}
\mathcal{P}_{2}: \underset{\mbx,\mby}{\textrm{minimize}}& \sum_{l=-(L)}^{L} \sum_{p=-P}^{P}\quad |r_{xy}^{l}(f_p)|^{2} \nonumber \\
\textrm{subject to} & \quad |x_k|=1 \; , |y_k|=1 \; , \forall \, k.
\end{align}\normalsize
The value of $P$ is governed by the maximum Doppler frequency of interest. The range of the interference causing the code shift $\hat{n}_{I}$ affects many range bins according to the relation $l=m-\hat{n}_{_I}$, therefore we choose a large enough $L$ in order to mitigate the effect of interference in all range bins.
 \begin{remark} \label{rem:1}
The expression in \eqref{eq:4} may be recast as 
 \begin{equation}\label{eq:r_xy}
r_{xy}^{l}(f_p)= \mbx^{\H}\Diag{\mbf_p}\mbC_l\mby
\end{equation}
where $\mbf_p=[1,e^{\j 2\pi f_p},\ldots,e^{\j 2\pi (K-1)f_p}]^{\T}$ and 
\begin{equation}
 \mathbf{C}_l = \mathbf{C}_{-l}^H= \begin{bmatrix} \bzero & \mbI_{K-l} \\ \mbI_l & \bzero \end{bmatrix}.
\end{equation}
 \end{remark}
The optimization problem \eqref{eq:opt6} is non-convex due to the unimodularity constraints. Herein, we propose to tackle the problem in a cyclic manner. Specifically, in the $s$-th iteration of our cyclic optimization algorithm , we first optimize $\mbx$ for fixed $\mby^{(s-1)}$ and set the optimal solution as $\mbx^{(s)}$. Then, ceteris paribus, we optimize $\mby$ for fixed $\mbx^{(s-1)}$. In the following, we present the solution to the two sub-problems involved in each iteration. But first, we remark on the unimodular quadratic programs (UQPs) and the power-method-like (PMLI) iterations to tackle such problems.
\begin{remark}\label{rem:2}
A UQP is defined as
\begin{equation}
\label{eq:UQP}
\underset{\mbx \in \Omega^{K}}{\textrm{maximize}} \quad\mbx^{\H} \mbG \mbx,
\end{equation}
where $\Omega^{K}=\{\mbx | x_k=e^{\j \omega}, \omega \in [0,2\pi), k \in \{0,\ldots,K-1\}\}$ is the set of unimodular vectors. The sequence of unimodular vectors at the $s$-th PMLI iteration
\begin{equation}
\label{eq:UQP_it}
\mbx^{(s+1)}=e^{\textrm{j}\arg{\mbG\mbx^{(s)}}},
\end{equation}
leads to a monotonically increasing objective value for the UQP, when $\mbG$ is a positive definite matrix. 
Moreover, in a UQP the diagonal loading technique is used to ensure the positive definiteness of the matrix, without changing the optimal solution. Particularly, in \eqref{eq:UQP}, the diagonal loading as
$\widetilde{\mbG}\leftarrow \lambda_{m} \mbI -\mbG$, with $\lambda_{_{m}}$ being slightly larger than the maximum eigenvalue of $\mbG$, results in an equivalent problem and leaves us with a positive definite $\widetilde{\mbG}$~\cite{soltanalian2014designing}.
\end{remark}
\noindent\textbf{ $\bullet$ Optimization of $\mbx$ for a fixed $\mby$:} By substituting~\eqref{eq:r_xy} in ~\eqref{eq:opt6}, the associated problem becomes 
\begin{align}
\mathcal{P}_{3}: \underset{\mbx}{\textrm{minimize}}& \quad \mbx^{\H} \mbB_y \mbx \nonumber\\
\textrm{subject to} & \quad |x_k|=1 \; , \forall \, k,
\end{align}
where $\mbB_y=\sum_{l=-(L)}^{L} \sum_{p=-P}^{P} \Diag{\mbf_p}\mbC_l \mby \mby^{\H} \mbC_l \Diag{\mbf_p}^{\H}$. By the diagonal loading technique introduced in Remark \ref{rem:2}, we have the positive 
definite matrix $\widetilde{\mbB}_y= \lambda_{m} \mbI -\mbB_y$ and obtain the equivalent problem
\begin{align}
\mathcal{P'}_{3}: \underset{\mbx}{\textrm{maximize}}& \quad \mbx^{\H} \widetilde{\mbB}_y \mbx \nonumber\\
\textrm{subject to} & \quad |x_k|=1 \; , \forall \, k.
\end{align}
\textbf{ $\bullet$ Optimization of $\mby$ for a fixed $\mbx$:} The problem with respect to $\mby$, mutatis mutandis, is 
\begin{align}
\mathcal{P}_{4}: \underset{\mby}{\textrm{maximize}}& \quad \mby^{\H} \widetilde{\mbB}_x \mby \nonumber\\
\textrm{subject to} & \quad |y_k|=1 \; , \forall k
\end{align}
where $\widetilde{\mbB}_x=\lambda_{m} \mbI -\mbB_x$ and $\mbB_x=\sum_{l=-L}^{L} \sum_{p=-P}^{P} \Diag{\mbf_p}\mbC_l \mbx \mbx^{\H} \mbC_l \Diag{\mbf_p}^{\H}$. 

We cyclically optimize the subproblems $\mathcal{P'}_3$ and $\mathcal{P}_4$, until convergence. Each of the subproblems is tackled by PMLI iterations introduced in Remark \ref{rem:2}. The steps of the proposed algorithm are summarized in Algorithm 1. The objective value of~\eqref{eq:opt6} at iteration $s$ is denoted by $J^{(s)}$. 
\begin{algorithm}
\caption{ PMCW waveform design for mutual interference mitigation}
 \label{algorithm_1}
 \begin{algorithmic}[1]
 \Statex \textbf{Initialize:} $\mbx^{0}$, $\mby^{(0)}$, $s=0$.
 \Statex \textbf{Output:} $\mbx^{*}$, $\mby^{*}$.
 \While {$|(J^{(s+1)}-J^{(s)})/J^{(s)}|\geq \epsilon$}
 \State Update $\widetilde{\mbB}_y^{(s)}$, $t\leftarrow 0$
 \Repeat \; $t \leftarrow t+1$
 \State$\mbx^{(s,t)}=e^{\j \arg{\widetilde{\mbB}_y^{(s)}\mbx^{(s,t-1)}}}$
 \Until{convergence}
 \State $ \mbx^{(s)}\leftarrow \mbx^{(s,t)}$
 \State Update $\widetilde{\mbB}_x^{(s)}$, $t\leftarrow 0$
 \Repeat \; $t \leftarrow t+1$
 \State$\mby^{(s,t)}=e^{\j \arg{\widetilde{\mbB}_x^{(s)}\mby^{(s,t-1)}}}$
 
 \Until{convergence}
 \State $ \mby^{(s)}\leftarrow \mby^{(s,t)}$
 \EndWhile
 \Statex \Return $\mbx^{*}=\mbx^{(s)}$ and $\mby^{*}=\mby^{(s)}$.
\end{algorithmic}
\end{algorithm}
In the following, we numerically evaluate the proposed algorithm.
\section{Numerical Evaluation}
\begin{figure}
\centering
\begin{subfigure}[b]{.49\linewidth}
\includegraphics[width=1\columnwidth]{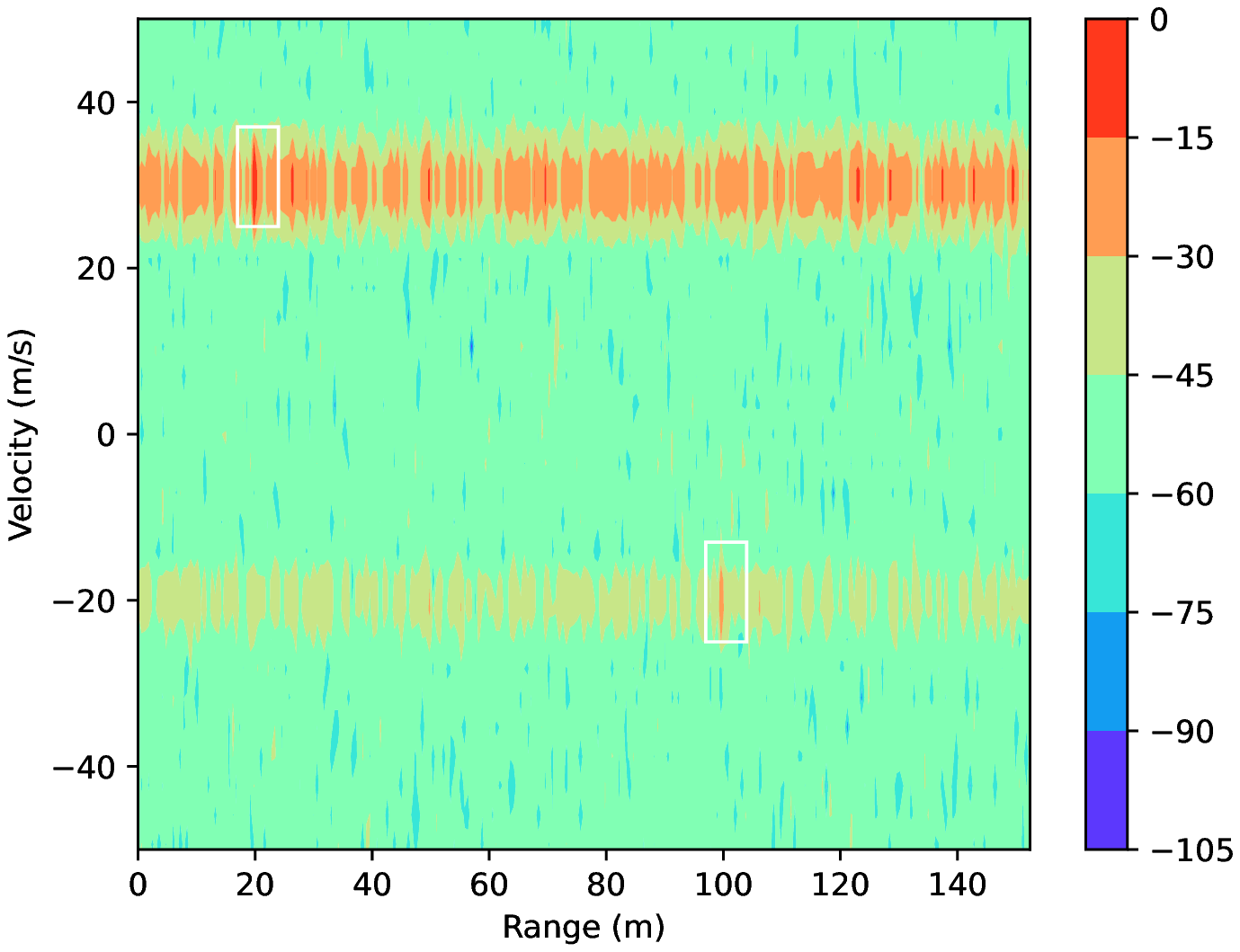}
\caption{}\setcounter{subfigure}{0}
\end{subfigure}
\begin{subfigure}[b]{.49\linewidth}
\includegraphics[width=1\columnwidth]{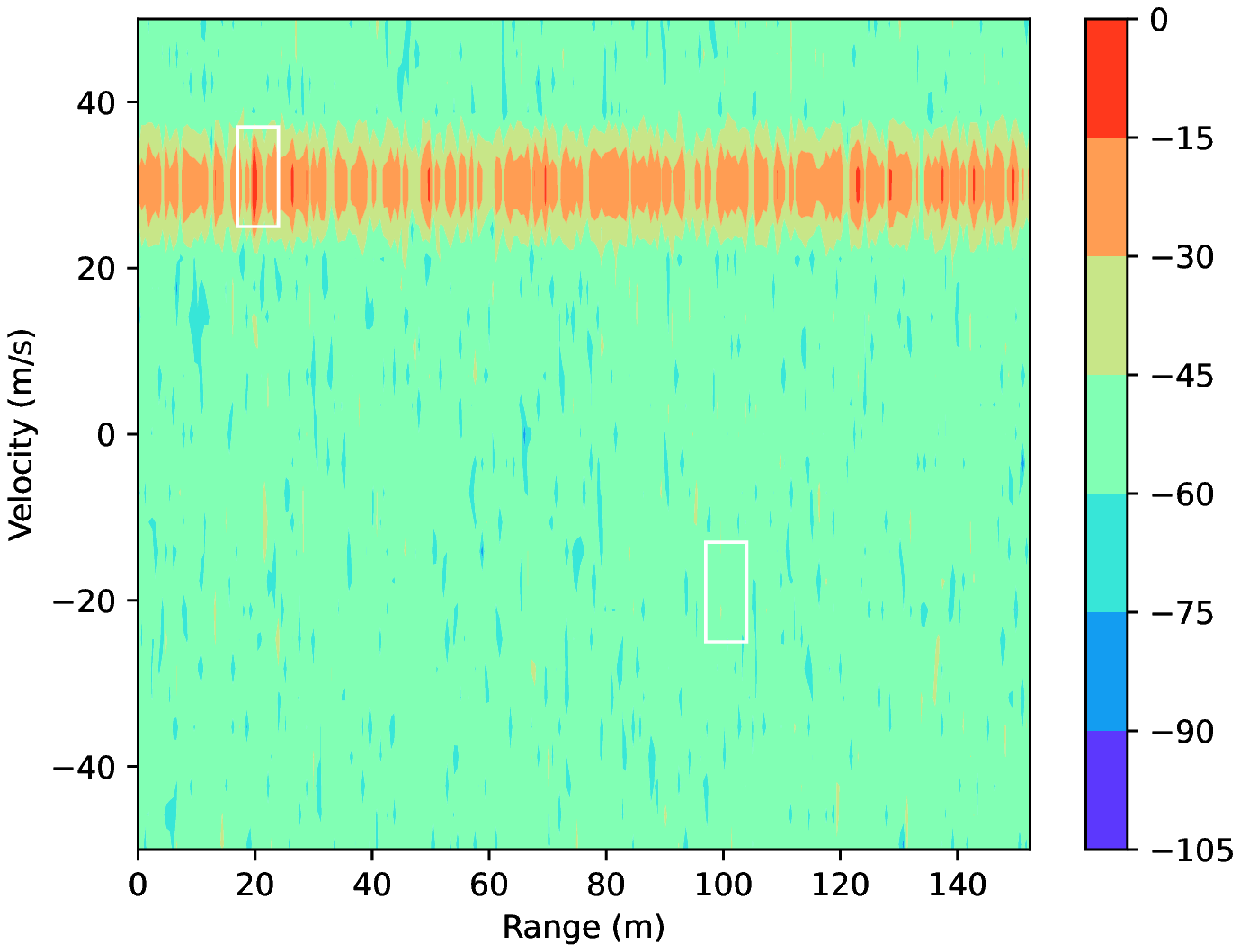}
\setcounter{subfigure}{1}
\caption{}
\end{subfigure}
\caption{\footnotesize Range-Doppler map of a single target with a) random PMCW signal and b) PMCW waveforms generated from Algorithm 1.
}
\label{fig:fig2}\end{figure}

We consider two vehicles mounted with radars transmitting PMCW waveform operating at $f_c=79$ GHz and pulse duration of $T=6.66$ ns. $N=140$  burst of the  signal is transmitted and a  white Gaussian noise distributed as $\mathcal{N}(\bzero,10^{-2}\mbI)$ is added to the received signal, and target and interferer RCS are assumed to be $35$ dBsm. The target is placed at a range $R=20$ m and moving with velocity $v=30$ m/s. The interferer radar is assumed to be located at a range $R_{_{I}}=200$ m with relative velocity $v_{I}=-20$ m/s. The 2D range-Doppler image of the target scene, as seen in the victim radar, is illustrated in Fig.~\ref{fig:fig2} (a) and (b) for PMCW waveform with $K=50$ chips generated randomly and using Algorithm~\ref{algorithm_1}, respectively. One can observe that in Fig.~\ref{fig:fig2}(a), the power
of the interference is strong which leads to false alarm.
The optimized PMCW waveforms appear to effectively mitigate the interference and therefore improve the target detection performance of the radar. 
\section {Discussion}
This paper examines the mutual interference between two PMCW radars and introduces a cost-effective computational algorithm for designing transmit waveforms based on unimodular quadratic programming. The proposed algorithm demonstrates excellent performance when the two radars cooperate and share the designed waveform. Extending this research to automotive systems with a significant number of MIMO radars, where minimizing interference between any pair is crucial, represents an ongoing and highly desired challenge.
\bibliographystyle{IEEEtran}
\bibliography{refs}
\end{document}